\documentclass[mathleft
]{an}
\usepackage{graphicx}
\usepackage{times}
\begin{document}

\Pagespan{1}{}
\Yearpublication{xxx}%
\Yearsubmission{2012}%
\Month{xx}%
\Volume{xxx}%
\Issue{xxx}%

\title{Hunting for binary Cepheids using lucky imaging technique}

\author{P. Klagyivik\inst{1}\fnmsep\thanks{Corresponding author:
  \email{klagyi@konkoly.hu}\newline}
\and  L. Szabados\inst{1}
\and  A. Szing\inst{1}
\and  Zs. Reg\'aly\inst{1}
}
\titlerunning{Hunting for binary Cepheids}
\authorrunning{P. Klagyivik et al.}
\institute{
Konkoly Observatory, Research Centre for Astronomy and Earth
   Sciences, Hungarian Academy of Sciences, H-1121 Budapest XII, 
   Konkoly Thege \'ut 15-17., Hungary
}

\received{xx yy 2012}
\accepted{xx yy 2012}
\publonline{later}


\abstract{
  Detecting companions to Cepheids is difficult. In most cases the
bright pulsator overshines the fainter secondary. Since Cepheids
play a key role in determining the cosmic distance scale it is
crucial to find binaries among the calibrating
stars of the period-luminosity relation. Here we present an
ongoing observing project of searching for faint and close companions
of Galactic Cepheids using lucky imaging technique.}

\maketitle

\section{Introduction}

\begin{figure}
\includegraphics[width=0.48\textwidth]{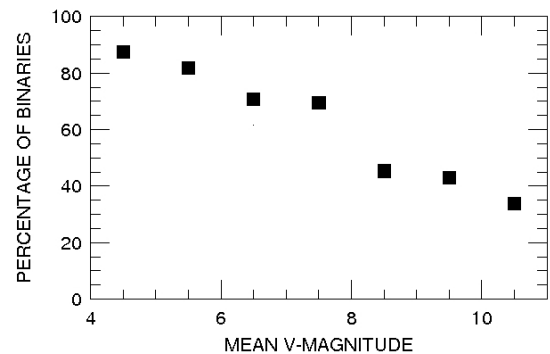}
\caption{Observed frequency of binaries among Galactic Cepheids
as a function of the mean apparent brightness (Szabados~et~al.~2011).
The trend seen in the diagram is the result of the observational selection effect.}
\label{binary_fraction}
\end{figure}

Cepheids are radially pulsating supergiant variable stars. They are located
in the instability strip on the Hertzsprung-Russell diagram.
According to stellar evolution theories stars between 4 and 20
solar masses cross this region during the He-burning phase in the core.
More than $50\%$ of the Galactic Cepheids
are members in binary or multiple systems (Sza\-bados~2003).
The fainter the target, the more difficult to
detect its even fainter companion(s). This is clearly seen
in the form of an observational selection effect in Fig.~\ref{binary_fraction}.

Undetected companions can significantly modify the relations
based on Cepheids. One of the most important is
the well known period-luminosity-color (P-L-C) relationship.
The binaries contribute to the dispersion of the relation
via the increase of the apparent brightness and the change of the observed color.
Therefore, the accuracy of distance determination is not at the point yet we
want to achieve. To improve the accuracy of the extragalactic distance
scale we have to omit these binaries from the calibrating sample.
For an overview of the importance of binary Cepheids see Szabados \& Klagyivik (2012).

From another point of view binary Cepheids play a very important role.
Cepheids in eclipsing binary systems are key objects for the determination
of stellar masses and for the solution of the long-standing
problem of Cepheid mass discrepancy (Christy~1968; Stobie~1969).
This discrepancy was mainly resolved with new opacity tables
(Moskalik~et~al.~1992), but there is still a $\mathbf{17\pm5\%}$ difference
between the masses determined from the evolutionary models and the pulsation
models (Keller~2008).
The list of Cepheids belonging to eclipsing binaries
in the Magellanic Clouds can be found in Szabados \& Neh\'ez~(2012).
In our Galaxy there is only one known
eclipsing binary containing a Cepheid component
(Sipahi~et~al.~2013).

There are many different ways to detect companions around Cepheids.
Here we present a survey of searching for close visual companions of
Galactic classical Cepheids using lucky imaging technique.

In Section 2 we overview the binary Cepheid searching methods,
in Sect. 3 we describe shortly the lucky imaging technique, in
Sect. 4 we present the details of our observing project. In the last part
we present the first results and we look into the practicalities
of this survey.

\section{Binary Cepheid searching methods}

The most effective way of binary detection is the spectroscopic investigation.
Since the pulsation and the orbital periods are independent
from each other the variability due to the two phenomena can be properly separated.

Binary detection using photometric methods are
based on the fact that if the companion has a different surface temperature
than the Cepheid, then the impact depends on the wavelength
of the observation. Klagyivik \& Sza\-ba\-dos (2009)
developed various methods to detect binary Cepheids based on
the peak-to-peak amplitudes observed in different photometric bands.
These methods are less effective than the radial velocity observations
but they can be used for faint stars as well.
If we are not able to correct for the photometric contribution
of the companion owing to the lack of brightness and color information
we have to exclude these Cepheids from the calibrations.

In the case of Cepheids bright outliers of the PL-relation
can be unrecognised binaries or blends (Fig. 6 in
Soszy\'nski et al. 2008).

Thanks to the Gaia astrometric satellite (to be launched
in October 2013) many stars will be marked expectedly
as astrometric binaries during the next decade.
The accuracy of the positions of Hipparcos ($10^{-3}$ arcsec)
was not able to resolve the orbits
of binary Cepheids, but the precision of Gaia ($10^{-6}$ arcsec)
will be sufficient in many cases to reach this goal.
Unresolved binaries however falsify the calculated parallaxes
as it was demonstrated by Szabados et~al. (2011).
All the negative parallax values of Hipparcos for Cepheids
within 2 kpc belong to known binary Cepheids.

Finding unknown visual (either optical or physical) binaries a very
good spatial resolution is needed. In some cases this is the only
way to discover the secondary source. High spatial resolution
optical imaging can be achieved in many ways. The main problem
is the turbulent behaviour of the terrestrial atmosphere.
Space telescopes are beyond this problem, the spatial resolution
depends only on the size of the mirror and the observing wavelength.
Adaptive optics are excellent tools to eliminate this adverse
atmospheric effect, but these instruments are too expensive for
smaller observatories. Lucky imaging is a good opportunity
to reach nearly diffraction limited spatial resolution
even from observing sites with average seeing.

Analysing $O-C$ diagrams, or radial velocity curves are also
viable methods to detect unknown companions. But these techniques
need gravitationally bound binary system and either short orbital
period or long term observations (e.g. Szabados et al. 2012).

Interferometry is also a plausible solution, it has better spatial resolution
than the lucky imaging technique, but it can be applied in the case of the brightest targets.

\section{Lucky imaging}

Long exposure time observations from the ground are always
affected by the atmospheric blurring.
Therefore the real spatial resolution is far from
the diffraction limited one.
Exposing thousands of images with very short exposure times
there will be some 'lucky' frames on which the random fluctuations
of the atmospheric turbulence are smaller and the image seems
stable in the small field of view of the camera.
The fraction of such frames depends on both the aperture
of the telescope and the seeing.
Tip-tilt correcting and adding the sharpest images much better spatial resolution
can be achieved than by using long exposure times.
In ideal cases the improvement of the image quality can reach a factor of $\sim 4$ (Smith et al. 2009).
The resulting point spread function is a diffuse halo
around the stars with a distinct, sharp core.

This technique, nowadays widely used by amateur astronomers observing planets,
is now used by professional astronomers, too.
For example in exoplanetary researches
it is a very interesting question how a companion star affects
the planet formation. Searching for close and faint
companions can help to answer this question
(Ginski et al. 2012, Bergfors et al. 2012).

The lucky imaging technique requires a usable fraction ($\sim 1-20\%$) of undistorted
images. Above 2.5 m in diameter this fraction is
effectively zero due to the size of the turbulent cells of the atmosphere.
More details of the technique can be found in Law et al. (2006).

Modern EMCCD (electron multiplying CCD) cameras are excellent instruments
for this technique (e.g. Smith et al. 2004). This technology allows very short readout of CCDs
with almost negligible noise.

\section{The observing project}

Although binary Cepheids are key objects in the calibration of the
cosmic distance scale we have insufficient information about
the companions of Cepheid type stars. In many cases we know only
the existence of the companion without additional information about
its physical properties (luminosity, temperature, etc.).

We launched an observing project at the Konkoly Observatory
searching for faint companions close to Galactic Cepheids
brighter than 13th mag in $I_C$ band using lucky imaging technique.
Our goal is to determine the fraction of visual binaries
among Cepheids and to measure the brightness and colors
of the companions.

\subsection{Observations}

The observations are obtained at the Piszk\'estet\H{o}
Mountain Station of the Konkoly Observatory using the
50 cm Cassegrain telescope.
The observatory is located 100 km far from Budapest to the northeast.
The seeing at an average night is 3-4$^{\prime\prime}$, but at an
excellent night it is near 1$^{\prime\prime}$.

The lucky imaging camera is an Andor iXon + 888 back-illuminated EMCCD camera.
This camera has 1024$\times$1024 active pixels with 13$\times$13 $\mu\rm{m}$ pixel size.
These parameters result in a $6.88 \times 6.88$ arcmin field of view
and a $0.40^{\prime\prime}$/px pixel scale.
The maximum readout rate is 10 MHz, while the frame rate is 8.9
frame per second at full frame without binning.
The peak quantum efficiency is $92.5 \%$ at 575 nm.
The EM gain multiplication can be varied linearly between 1 and 1000.
Using this EM gain multiplication the readout noise is less than $1 e^{-}$.

During the observations we used the camera in 14 bit mode with 10 MHz horizontal readout speed.
The EM gain varied between 100 and 300. The exposure times were between 10-40 ms.
For each star we obtained 15000 - 30000 exposures mostly in the $I_C$ band.

The diffraction limit of the telescope is $0.40^{\prime\prime}$ in $I_C$ band,
$0.33^{\prime\prime}$ in $R_C$ and $0.27^{\prime\prime}$ in $V$ band, but
at shorter wavelength the seeing is much worse. So in $V$ and $R_C$
bands the pixel scale of the camera is larger than the diffraction limit.
Therefore if we want to reach nearly diffraction limited spatial resolution the $I_C$
band is the appropriate filter for lucky imaging observations.
Comparing the results we sometimes observed in other filters, too.

We plan to use the 1m RCC telescope also located at the Piszk\'estet\H{o}
Mountain Station. Suspected binaries are planned to be confirmed
with an even larger telescope.

\section{First results}

So far we have observed 76 northern Galactic Cepheids of our 120
program stars. For this technique the best targets are near the zenith.
The zenith distance of our program stars at culmination should not exceed
$30^{\circ}$, which corresponds to a declination range of
$18^{\circ} < \delta < 78^{\circ}$. The brightness of the targets is between
$7$th and $13$th mag in $I$ band. The data reduction pipeline is in the test phase.

\subsection{CE Cas}

CE Cas is a unique object in our target list. It is located in the NGC 7790
open cluster. Both components of this binary system are Cepheids.
The pulsation periods of the components are 5.141 (CE Cas A)
and 4.479 days (CE Cas B).
The apparent mean visual magnitudes of the components are
$10.90$ and $11.02$ mag, respectively in $V$ band.
The separation is $2.5^{\prime\prime}$, so on the nights of good seeing
they can be separated even in long exposure time images.

\begin{figure*}
 \centering
 \includegraphics[width=0.96\textwidth]{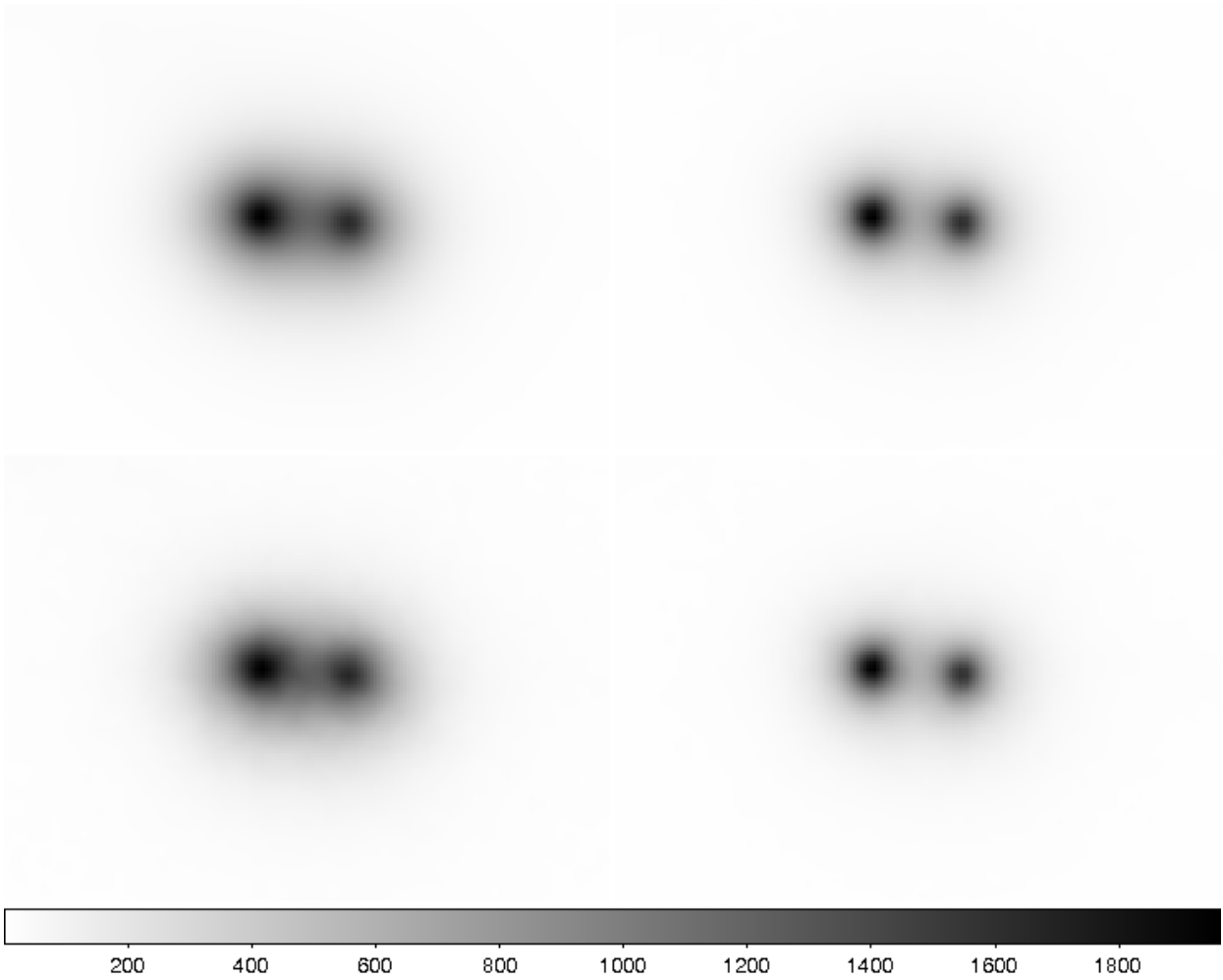}
\caption{Comparison of CE Cas with (right) and without (left) lucky imaging technique.
Top left panel: sum of all the images only with tip-tilt correction,
bottom left panel: sum of 300 randomly selected images,
top right panel: sum of the best 750 ($5\%$) frames,
bottom right panel: sum of the best 300 ($2\%$) images.}
\label{CECas_lucky}
\end{figure*}

On 4 October 2011 we obtained 15000 images in $V$, $R_C$ and $I_C$ bands
with 30 ms, 30 ms and 40 ms exposure times, respectively.
The seeing was $3^{\prime\prime}$.
After the standard image reduction steps -- bias, dark and flat corrections --
we selected differential number of images, shifted them to each other and added.
Each pixel was upscaled to $4 \times 4$ pixels.
The result is shown on the right side of Fig.~\ref{CECas_lucky}.
In the top left panel we just summarized all the images only with tip-tilt correction
and without any selection. In the top right panel we selected the best 750 ($5\%$)
of the frames. In the bottom panels we added 300 ($2\%$) images, but
on the right side the best images were taken, while putting in context
in the left side we selected the frames randomly. The pictures clearly
show that the selection is working efficiently. Random selection results in the same
image as no selection (left panels). There is only marginal difference
between the $5\%$ and the $2\%$ selection. But using the best $10 \%$ the
result is significantly worse.

\begin{table}[h!]
\caption{Average FWHM values in the field of CE Cas in $V$, $R_C$ and $I_C$ bands
after tip-tilt correction with and without lucky imaging. The improvement ratios
were calculated as the ratio of the FWHM without and with lucky imaging.
During the lucky imaging the best 300 frames were selected.
}
\label{fwhm}
\begin{tabular}{cccc}\hline
 & \multicolumn{2}{c}{FWHM ($^{\prime\prime}$)}&\\
\cline{2-3}
Band & no lucky & with lucky & improvement ratio\\
\hline
$V$   & $1.32$ & $1.14$ & $1.16$\\
$R_C$ & $1.54$ & $1.15$ & $1.34$\\
$I_C$ & $1.68$ & $1.11$ & $1.51$\\
\hline
\end{tabular}
\end{table}

The spatial resolution increased by a factor of $1.5$ in $I_C$ band (Table \ref{fwhm} and Table \ref{fwhm_I}),
while the brightness of the central peak of the stars is doubled. The data reduction
pipeline is under development, so there is still potential
to improve the spatial resolution. The pipeline will be published soon
(Reg\'aly, in prep.).

We found no more optical stellar companion in the system of CE Cas.

Since the observations and data reductions are ongoing,
calculation of limits of separation and magnitude difference for the detectability of
companions is left for the future.

\begin{table}[h!]
\caption{Tip-tilt corrected FWHM values in $I_C$ band for different number
of selected images.
}
\label{fwhm_I}
\begin{tabular}{cccc}\hline
 \noalign{\smallskip}
$\%$ & FWHM ($^{\prime\prime}$)\\
\hline
$100 \%$& $1.68$\\
$10 \%$ & $1.53$\\
$5 \%$  & $1.17$\\
$2 \%$  & $1.11$\\
\hline
\end{tabular}
\end{table}

\section{Conclusion}

Lucky imaging is a powerful tool detecting faint and/or close
companions around bright stars which are not detectable with
other techniques. The period-luminosity relation of Cepheids
is one of the most important relations in the field of the
cosmic distance scale, but binaries disturb the calibration.
It is indispensable to exclude these systems even if the
companion is faint.

The first results with the 50 cm Cassegrain telescope at Piszk\'estet\H{o}
Mountain Station and the EMCCD camera are promising.
On an average night with $3^{\prime\prime}$ seeing we
reached almost three times better spatial resolution
than using long exposure time images.
On much better nights we should reach nearly the diffraction
limited resolution.

\acknowledgements

This project has been supported by ESTEC Contract No. 4000106398/12/NL/KML.
Peter Klagyivik acknowledges support from the Hungarian State E\"otv\"os Fellowship.

\end{document}